\documentclass{aastex}

\usepackage{apjfonts}
\usepackage{emulateapj5}

\newcommand{\mkfiga}[2]{\noindent \parbox[t]{0.48\textwidth}{
   \plotone{#1} \figcaption{#2} \vspace*{2.5em} \noindent}}
\newcommand{\mkfigc}[2]{\noindent \parbox[t]{0.48\textwidth}{
   \plotone{#1} \figcaption{#2} \noindent}}

\newcommand{\bvfreq}{Brunt-V\"ais\"al\"a frequency}

\newcommand{\Mstar}{{ M}_\star}
\newcommand{\he}[1]{\mbox{$^{#1}$}{\rm He}}

\begin{document}

\submitted{Received 2000 August 28; accepted 2000 November 16}


\title{The effect of \he{3} diffusion on the pulsational spectra \\
of DBV models}

\author{M. H. Montgomery}
\affil{
Institute of Astronomy, University of Cambridge, Madingley Road, 
Cambridge CB3 0HA, United Kingdom
}
\email{mikemon@ast.cam.ac.uk}
\and
\author{T. S. Metcalfe \& D. E. Winget}
\affil{
{McDonald Observatory and Department of Astronomy,
The University of Texas, \\ Austin, TX 78712, USA.} }
\email{travis@astro.as.utexas.edu dew@bullwinkle.as.utexas.edu}

\begin{abstract}

We consider envelopes of DB white dwarfs which are not composed of pure
$^4$He, but rather a mixture of $^{3}$He and $^4$He. Given this
assumption, the same diffusive processes which produce a relatively
pure H layer overlying a He layer in the DA's should work to produce a
$^{3}$He layer overlying a $^4$He layer in the DB's.  We examine the
relevant timescales for diffusion in these objects, and compare them to
the relevant evolutionary timescales in the context of the DBV white
dwarfs.  We then explore the consequences which \he{3} separation has
on the pulsational spectra of DBV models. Since GD~358 is the
best-studied member of this class of variables, we examine fits to its
observed pulsation spectrum. We find that the inclusion of a \he{3}
layer results in a modest improvement in a direct fit to the periods,
while a fit to the period {\em spacings} is significantly improved.

\end{abstract}


\section{Astrophysical Context}

The two greatest successes of white dwarf asteroseismology have come
from the analysis and interpretation of Whole Earth Telescope (WET)
data on the objects GD~358 and PG~1159 \citep{Winget91,Winget94}. The
temporal spectra of both objects show well-defined multiplet structures
as well as many consecutive radial orders of the same $\ell$. This
allows us immediately to estimate the rotational frequency, $\Omega$,
and the mean period spacing, $\langle \Delta P \rangle_{\ell}$, with the
latter quantity putting strong constraints on the mass of the best-fitting
model. Furthermore, the variations from equidistant period spacing, i.e.,
``mode trapping'', give us information (at least in the case of GD~358)
about the radial structure of the star, such as the thickness of its
surface He layer.

In a detailed analysis of the frequency distribution in GD~358,
\citet{Bradley94a} found evidence of a chemical transition zone located
at a mass depth of $1.5 \times 10^{-6} \Mstar$, which they interpreted
as the C/He boundary. A He layer thickness of $\sim$10$^{-6} \Mstar$ is
thinner than expected on naive evolutionary grounds, as well as thinner
than suggested for the DA white dwarfs by the asteroseismological work
of \citet{Clemens93}.  While the fit of Bradley \& Winget explained
the trapping features in the neighborhood of 700 sec reasonably well,
it did not do as well for periods near 500 sec. They found evidence that
these periods near 500 sec might be better fit if an additional chemical
transition zone at a depth of $\sim$10$^{-2} M_{\star}$ is assumed.

Independent of the above considerations, if the theory of diffusion of
chemical species in stars is correct, then at least a partial separation
of different isotopes of the same element must occur in white dwarfs, as
was pointed out to one of us by Clayton (1988, private communication).
At the time, there were no known observational consequences; this has
since changed with the advent of white dwarf asteroseismology. 

While the Galactic number ratio of \he{3} to \he{4} has been measured
to be of order $10^{-4}$ in many astrophysical environments
\citep{Galli95,Prantzos96}, this does not necessarily hold for the
stellar cores which are presumably the white dwarf progenitors. In
fact, standard evolutionary theory would suggest quite low levels of
\he{3} in these objects \citep{Galli95}. Our view is that given the
theoretical uncertainties in the late stages of stellar evolution, we
should not rule out any possibilities, and indeed should seek to make
as many independent measurements as possible.

These considerations led to two independent motivations for considering
\he{3} diffusion \citep{Winget98}.  First, it is a process which may be
a generic feature of white dwarf cooling, and, as such, should be
considered as part of our asteroseismological analyses.  Second, it
might allow us to place the mode trapping data for GD~358 in a
different context: this star could have a C/\he{4} transition zone at
$\sim 10^{-2} \Mstar$ and a thinner \he{4}/\he{3} transition zone at
$\sim 10^{-6} \Mstar$. If this is the case, then the DBV's and the
DAV's would again have the same order of magnitude He layer
thicknesses. To determine the plausibility of these hypotheses, we
first examine the relevant diffusion timescales, and then the effect
which such a layering structure has on our fits of GD~358.

\section{Diffusion}
\subsection{Timescales}

The process of diffusion is of major importance, since we are aware of
no other effect which could lead to the spatial separation of \he{3} and
\he{4}. We therefore examine the relevant timescales for such diffusive
processes, to see if it is plausible for a significant fraction of the
\he{3} to have separated from \he{4} in the elapsed evolutionary times
for these objects.

Fortunately for us, \citet{Fontaine79} have already examined
the related problem of C diffusion in a background of normal He, i.e.,
\he{4}. In their analysis, they treated C as a trace element. This is
an excellent approximation for our case as well, since we naively expect
that the \he{3} is only about one part in $10^{4}$ of the \he{4}. 

\mkfiga{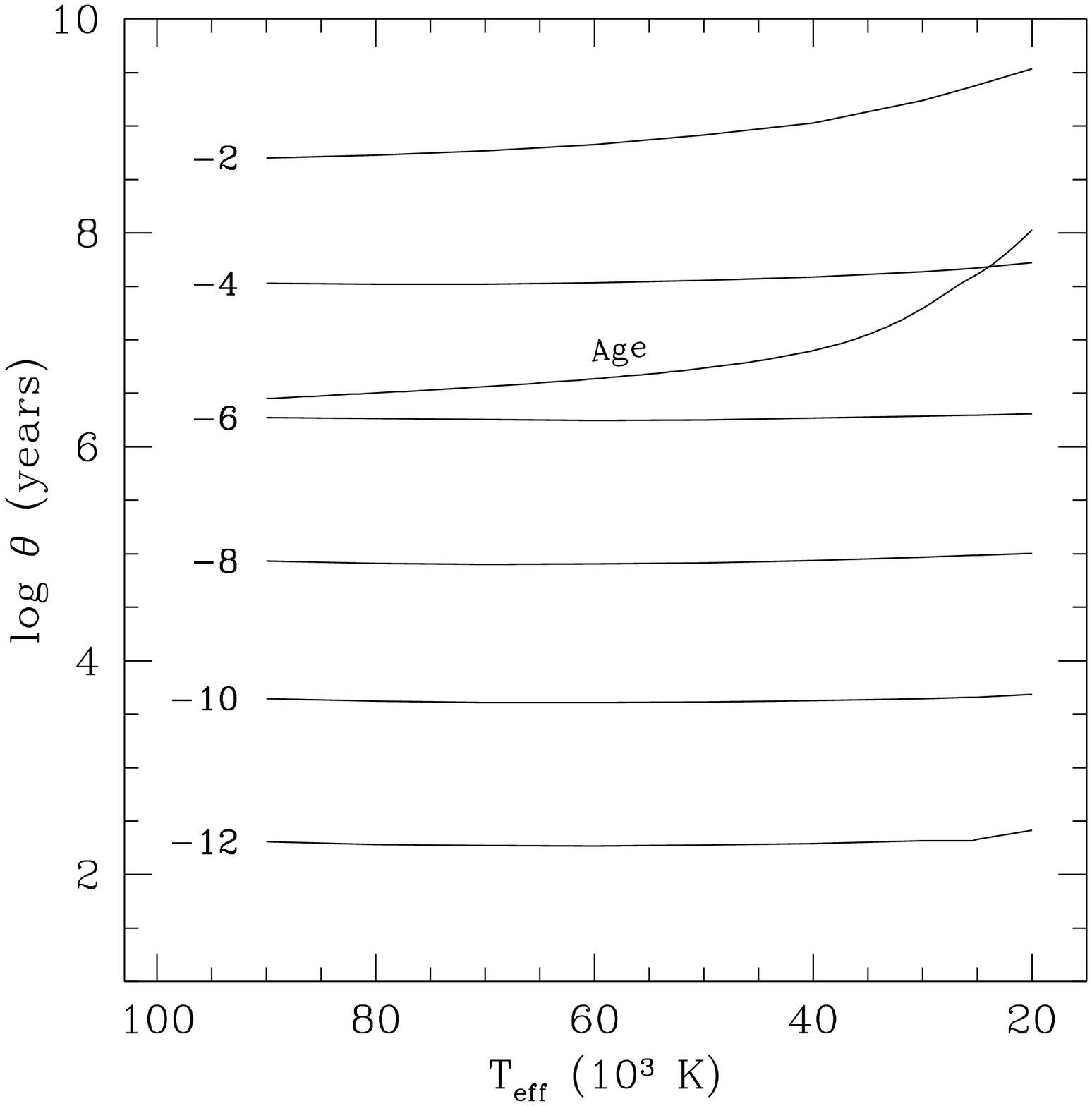}{
The timescale, $\theta$, for diffusion of \he{3} in a background of
\he{4} as a function of $T_{\rm eff}$ at different mass depths between
-12 and -2. The curve labeled ``Age'' gives the age of the model as a
function of $T_{\rm eff}$; below the line the profile should have reached
diffusive equilibrium and above it little diffusion should have occurred.
\label{diffuse}
}
\vspace*{0.4em}

Applying their equations for this case for a 0.612 $M_{\odot}$ model,
we obtain the result shown in Figure~\ref{diffuse}.  We see that if
we assume a $T_{\rm eff}$ of 25,000~K for a typical DBV, then diffusive
equilibrium between \he{3} and \he{4} should prevail down to approximately
the $10^{-4} M_{\star}$ mass point. Thus, if the \he{4} layer is as thick
or thicker than this, and the \he{3} abundance is $\sim 10^{-4}$ that
of the \he{4} abundance, we would expect a \he{3} layer of thickness
$\sim 10^{-8}$.

One clear prediction of this is that the spectra of DB's in this
temperature range should show more or less pure \he{3}, at least for
\he{4} layers which are thicker than $\sim 10^{-6}$. This is currently in
the process of being tested using high-resolution spectroscopic
observations (Koester, private communication).

We mention as an aside that the case of \he{3} may be unique in that it is
a trace isotope which is {\em lighter} than the dominant species, so that
it can diffuse upwards and produce a thin, but not asteroseismologically
negligible layer. For trace isotopes which sink, there should be no
such signature.

\vspace*{1.0em}
\subsection{Profiles}

In this preliminary exploration, we will only consider \he{4}/\he{3}
zones with equilibrium profiles, although the profiles could in fact
be less sharp than this.  The treatment of the chemical profiles in
the transition zones in our present and previous work is based on
the work of \citet{Arcoragi80}.  Essentially, we use equation (A6)
of \citet{Arcoragi80}, which assumes that an equilibrium distribution
has been reached and that one of the elements may be treated as a trace
element. This is certainly a valid assumption for the initial stages of
\he{3} diffusion, given the expected abundance ratio of \he{3} to \he{4}.

\section{Equation of State}

To include the effects of a \he{3} layer we have made a relatively simple
modification to the envelope routines in our evolutionary code. Since
we are interested in modeling a DB, we are free to use the array space
normally reserved for the H profile and use it for the \he{3} profile.
To this end, we have replaced the H equation of state (EOS) and opacities
with those appropriate for \he{3}. In doing this, we have taken the \he{3}
EOS and opacities to be equal to those of \he{4} at 3/4 the density
(to correct for the isotopic mass ratio). This approximation should
be more than sufficient for our purposes.  A comparison of the region
of period formation for the cases of a \he{4}/H and a \he{4}/\he{3}
envelope is given in \citet{Montgomery99d}.

\section{Previous Best Fits}

We now wish to examine the effect which a \he{3} layer could
have on the pulsation frequencies. First, in the top panel of
Figure~\ref{fit0.61}, we show one of the best-fit models for the star
GD~358 from \citet{Bradley94a}.  The filled circles connected by solid
lines show the observed mode trapping structure, and the open circles
connected by dotted lines show the results from the best-fit model. This
model has $M_{\star}/M_{\odot} = 0.61$, $T_{\rm eff} =$ 24,044 K, and
$M_{\rm 4He}/M_{\star}=1.5\cdot 10^{-6}$. The lower 3 panels show the
effect a thin layer of \he{3} has on the mode trapping structure. If
the layer is as thick as $M_{\rm 3He}/M_{\star}=1.5\cdot 10^{-10}$,
corresponding to $N(\he{3})/N(\he{4})=10^{-4}$, then we see that the mode
trapping structure in the vicinity of 700 sec is significantly altered.
\mkfigc{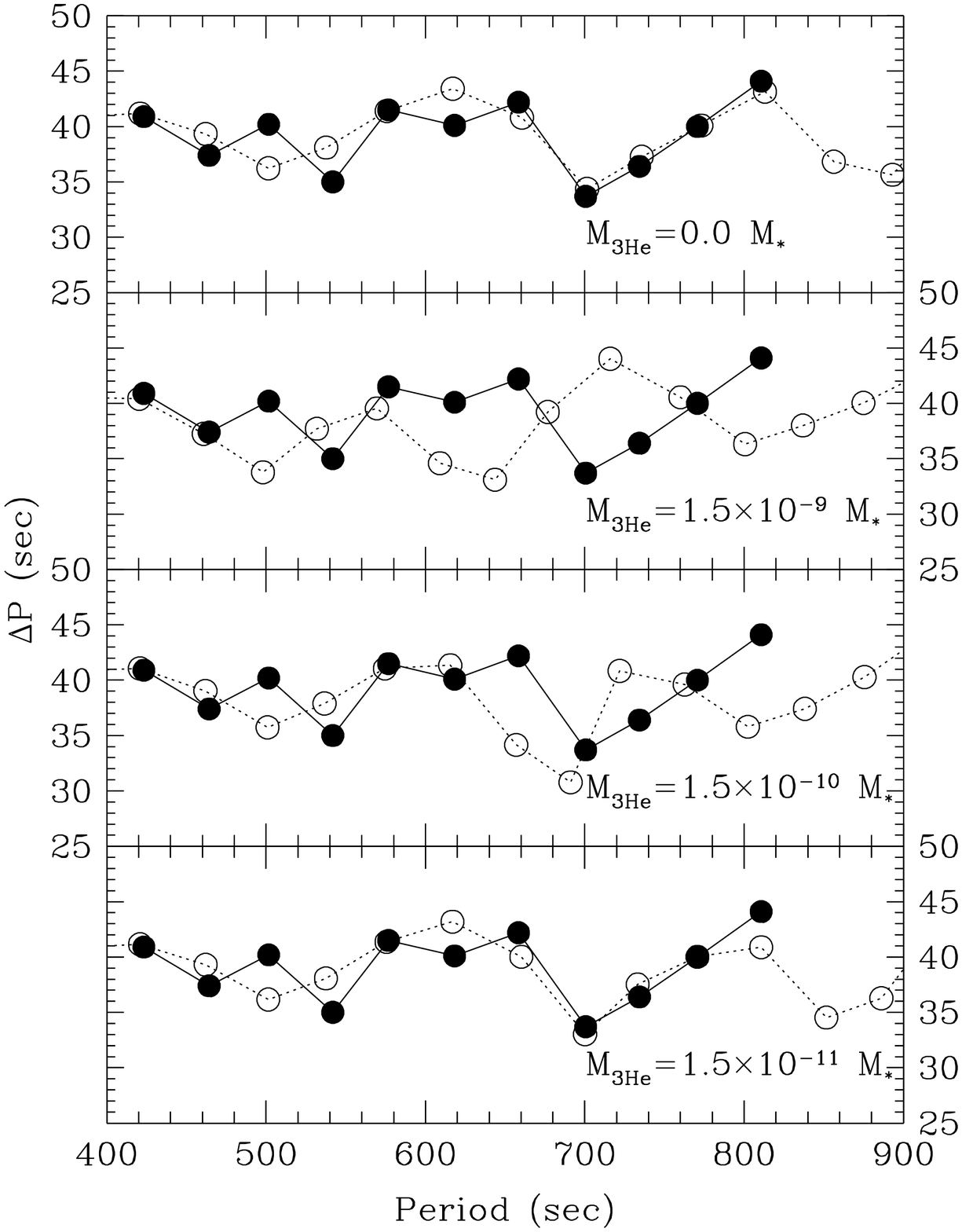}{
A comparison of the observed trapping (filled circles, solid
lines) with carbon core models (open circles, dotted lines). The top
panel is the best-fit model of \citet{Bradley94a}, which of course has
\mbox{$M_{\rm 3He}=0.0$}. The lower panels show the mode trapping structure
for nonzero values of $M_{\rm 3He}$, as indicated in each panel.
\label{fit0.61}
}
\begin{table*}
\tabcaption{Results of Fits using a Genetic Algorithm
\protect\label{results}
}
\vspace*{-0.4em}
\begin{center}
\begin{tabular}{lcccccc}
\hline\\[-0.95em]
\hline\\[-0.7em]
\multicolumn{1}{c}{Model} & $T_{\rm eff}$ (K) & $M_{\star}/M_{\odot}$ & $\log M_{\rm 4He}/M_{\star}$ &
$\log M_{\rm 3He}/M_{\star}$ & $\sigma(P)$ & $\sigma(\Delta P)$ \\[0.3em]
\hline\\[-0.7em]
C/thick4He     & 23,100  & 0.655  & -2.74  & \nodata &   2.30 & 2.65 \\
C/thick4He/3He & 23,200  & 0.605  & -2.74  &  -6.02  &   1.99 & 2.29 \\[0.7em]

C/thin4He      & 23,800  & 0.600  & -5.76  & \nodata &   2.71 & 2.53 \\

C/thin4He/3He  & 24,400  & 0.580  & -5.71  & -10.09  &   1.95 & 2.66 \\[0.7em]

O/C/4He\tablenotemark{a} & 22,600 & 0.650 & -2.74 & \nodata &  1.50 & 1.92 \\
O/C/4He/3He\tablenotemark{a} & 22,300 & 0.630 & -2.79 & -7.49 & 1.29 & 1.32 \\[1.0em]

Inferred Errors &  ~~~~500   & 0.015  &  ~0.06  &  ~0.12   &  \nodata & \nodata \\[0.3em]
\hline
\end{tabular}
\end{center}
{\footnotesize \hspace{19ex} $^a$These fits were performed using $f =
1/[\sigma(P)+\sigma(\Delta P)]$ as the fitness criterion; for the
other models $f = 1/\sigma(P)$ was used.}
\end{table*}

 \noindent
Unfortunately, this is the region in which the fit was already quite good,
whereas in the region around 500 sec, where the fit was poor, there is
very little change. However, this analysis clearly demonstrates that
the inclusion of a \he{3} layer can have a measurable effect on the
calculated mode trapping structure.

\section{New Fits using a Genetic Algorithm}

In an effort to test asteroseismologically the viability of the hypothesis
of a \he{3} layer, we performed an extensive set of calculations. Using
a genetic algorithm \citep{Metcalfe99,Metcalfe00}, the parameters
$M_{\star}$, $T_{\rm eff}$, $M_{\rm 4He}$, and $M_{\rm 3He}$ were varied
to produce pulsational models whose $\ell=1$, $m=0$ periods were compared
with those inferred from the observations. The region of parameter space
explored is given by:
\begin{displaymath}
\begin{array}{ccccc}
0.45~M_{\odot} & < & M_{\star} & < & 0.95~M_{\odot} \\
20,000 {\rm~K} & < & T_{\rm eff} & < & 30,000 {\rm~K} \\ 
\sim 10^{-7}~M_{\star} & < & M_{\rm 4He} & < & 10^{-2}~M_{\star}\\
10^{-5}~M_{\rm 4He} & < & M_{\rm 3He} & < & 10^{-3}~M_{\rm 4He}
\end{array}
\end{displaymath}
The best-fit models from these runs are listed in Table~\ref{results}.
The column labeled $\sigma(P)$ gives the residuals (standard deviation)
of the calculated and observed periods, and the column labeled
$\sigma(\Delta P)$ gives the residuals of the calculated and observed
period {\em spacings}. For the upper four models we have taken the
fitness criterion to be $1/\sigma(P)$, whereas for the lower two models
we took it to be $1/[\sigma(P)+\sigma(\Delta P)]$; thus, for these last
two entries, we are fitting not just the periods but the spacings
between consecutive periods. We have chosen this criterion for the
models with a more realistic core composition based on our suspicion
that the period {\em spacings} will be a more sensitive diagnostic of
the \he{3} layer than the periods themselves will be.

As a reference to previous fits, the fit of \citet{Bradley94a} for a C
core model has $\sigma(P) \sim 2.3$~sec. Thus, the present best fits,
both with and without a \he{3} layer, having period residuals of $\sim
1.5$ and $1.3$~sec, respectively, represent a significant improvement
over previous fits.

\section{Discussion}

We see from Table~\ref{results} that the fits with O/C cores are
significantly better than those with pure carbon cores, independent of
whether a \he{3} layer is present or not, a result previously found by
\citet{Metcalfe00}. In Figure~\ref{fit}, we display the mode trapping
diagrams for these two fits. Both fits reproduce the periods quite
well, with the \he{3} model reproducing the period {\em spacings} much
better ($\sim 1.3$ sec compared to $\sim 1.9$ sec). This indicates that
the period spacing may be a better diagnostic for the fine structure
produced by the \he{4}/\he{3} transition zone than just the periods
themselves.

We now seek to understand the relative importance of the O/C and
\he{4}/\he{3} transition zones. In the asymptotic limit of high radial
overtones and large periods, the frequency of a given mode is given by
a simple radial integral of the \bvfreq. Using a ``period formation''
diagram, we can show the relative weight which a given region has in
determining a mode's period.

In Figure~\ref{pform2}, we show such a diagram for the case of our
best-fit \he{3} model.  The three peaks which are labeled correspond to
the O/C, C/\he{4}, and \he{4}/\he{3} transition zones. Using the
equilibrium diffusion coefficients for the \he{4}/\he{3} 
transition zone, we see that it is of relatively minor importance in
determining the mode frequencies. In contrast, the O/C and C/\he{4}
\mkfigc{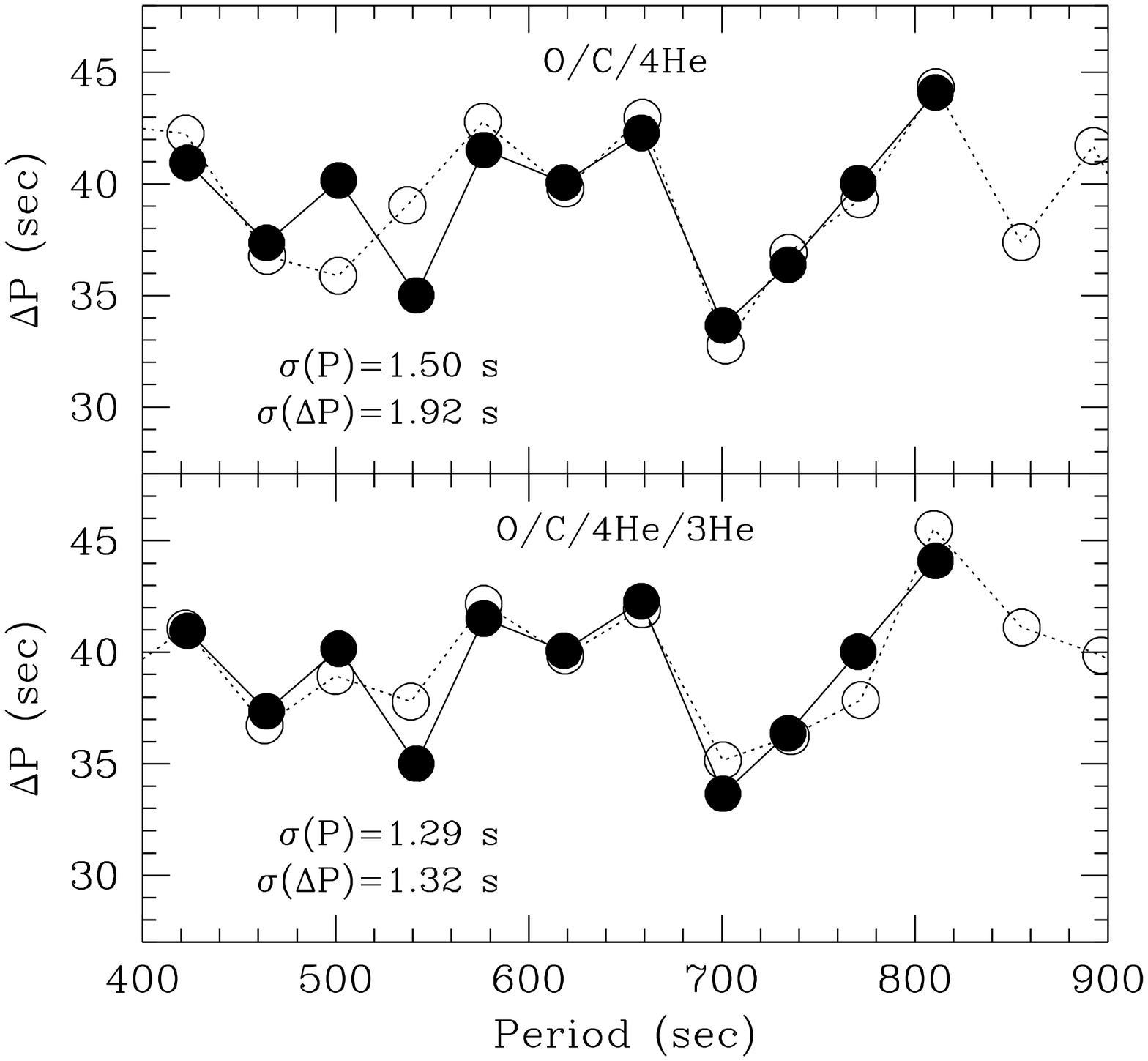}{
The upper panel shows the best-fit model without \he{3},
and the lower panel the best-fit model with \he{3}. As can be seen,
these fits have much smaller residuals than the previous fit given in
the upper panel of Figure~\protect\ref{fit0.61}.
\label{fit}
}
\mkfiga{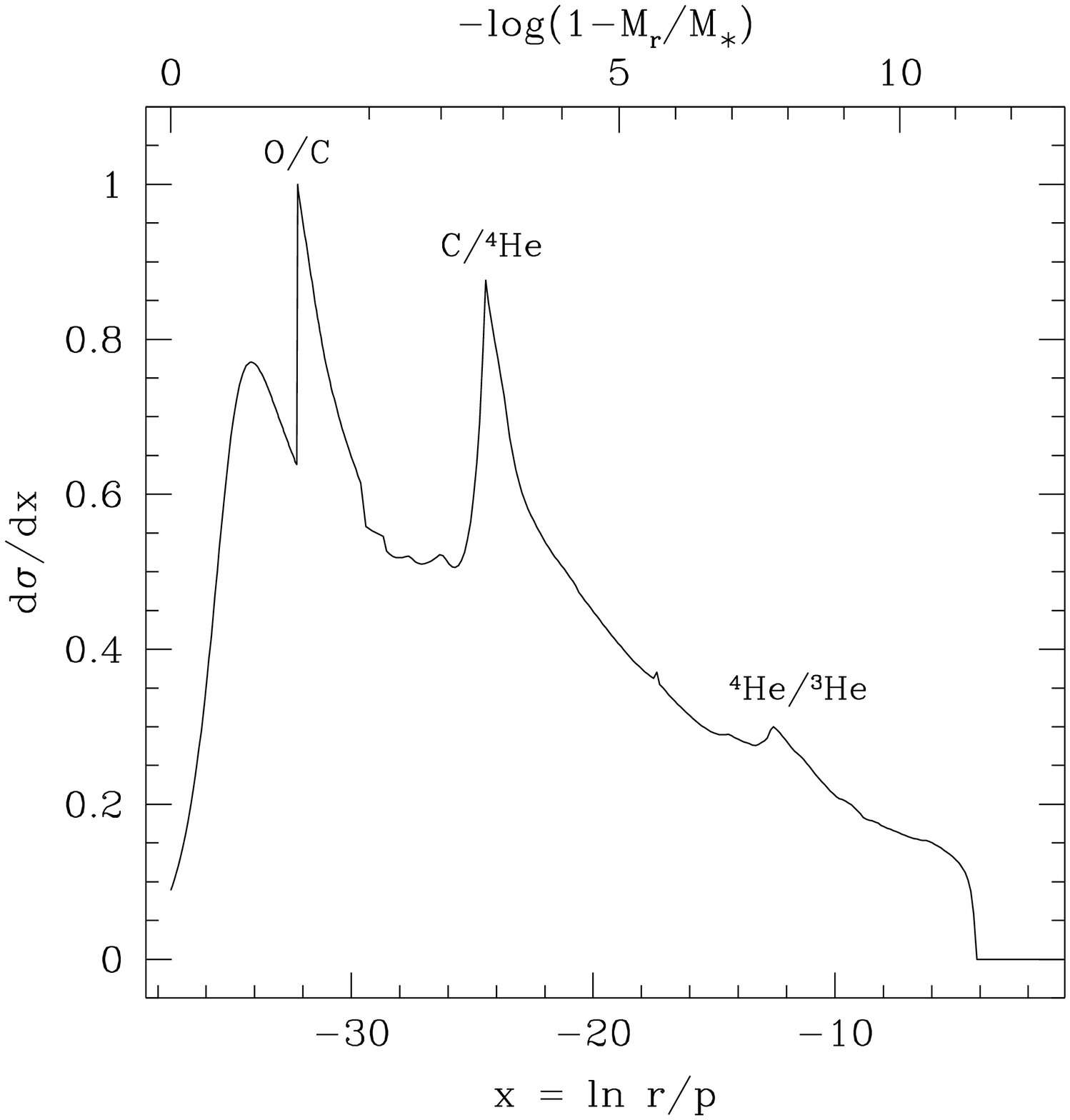}{
The relative contribution of a region to the
frequency of a mode, $\sigma$, as a function of $x = \ln (r/p)$,
where $r$ is the radius and $p$ is the pressure, both in cgs units.
The features corresponding to the O/C, C/\he{4}, and \he{4}/\he{3}
transition zones have been labeled.
\label{pform2}
}
transition zones are both quite pronounced and should significantly
affect the periods of the modes calculated in the models; this is borne
out by the major improvement in the standard deviation of the periods,
$\sigma(P)$, with the inclusion of an O/C chemical profile. The period
{\em spacings}, on the other hand, are a differential quantity, and
therefore sensitive to even small deviations in the background
structure. It is therefore not surprising that the inclusion of a
\he{3} layer results in a major reduction in the resduals of the period
spacings.

We now attempt to quantify the statistical significance of the improvement
of the fit when the parameter corresponding to the \he{3} layer thickness
is added. Following \citet{Koen00}, we apply the Bayes Information
Criterion (BIC) for $N=11$ data points. We find that the addition of
a parameter should be accompanied by a decrease in the residuals of at
least $\sim$10\% in order to be considered statistically significant.
We note that this is a necessary but certainly not sufficient condition
for the validity of adding a parameter.

Examining the best-fit O/C core models, we see that the standard
deviation of the fit to the period spacings improved from 1.92~sec to
1.32~sec with the addition of \he{3}. This is a decrease of 30\% and
should therefore be considered statistically significant.

Finally, we again note that the models with O/C cores yield markedly
lower residuals, regardless of which criterion is used (periods or
period {\em spacings}). This result has recently been obtained by
\citet{Metcalfe00}, who report the results of such fits in an extensive
parameter space of white dwarf models.

\section{Conclusions}

From this preliminary analysis, we have shown that diffusion theory
applied to white dwarfs predicts that any \he{3} initially present in
the DBV's down to a depth of $\sim$10$^{-4} \Mstar$ should have diffused
upward to produce a surface layer of \he{3}. In addition to being detectable
spectroscopically (Koester, private communication), such a \he{3} layer
can significantly affect the asteroseismological fits,  and therefore
needs to be included in such analyses.

We find that although the inclusion of a \he{3} layer results in only a
marginal improvement to the fits to the periods, the fit to the period
{\em spacings} is significantly improved. This is because the period
spacings are more sensitive to the fine structure which a \he{3}/\he{4}
transition zone produces than the periods themselves are.

Finally, we find that an O/C core (essentially, the transition in
the core from an O/C mixture to pure C) fits the observed pulsational
spectrum of GD~358 {\em much} better than a pure C core, in agreement
with \citet{Metcalfe00}, who first obtained this result. This gives
us the hope of someday being able to constrain the prior nuclear
burning history of GD~358 and other pulsating white dwarfs.

\acknowledgements

We would like to thank Jim Truran, Don Clayton, and Hugh Van Horn for
useful discussions on this topic. This work was supported in part by
the Austrian Fonds zur F\"orderung der wissenshaftlichen Forschung,
project number S7304, by the National Science Foundation under grant
AST-9876730, and by the National Aeronautics and Space Administration
under grant NAG5-9321.


\end{document}